\documentclass[useAMS,usenatbib,8pt]{mn2e}
\usepackage{hyperref}
\usepackage{graphicx}
\usepackage{subfig}

\usepackage{amssymb,psfrag}

\title[Constraints on power-law cosmology
]{Observational constraints on Hubble constant and deceleration parameter in power-law cosmology}
\author[\textbf{S. Kumar}]{ \textbf{Suresh Kumar} \thanks{E-mail: sukuyd@gmail.com; sureshk@iucaa.ernet.in}\\
\noindent Department of Applied Mathematics, Delhi Technological University, Bawana Road, Delhi 110 042, India
}

\begin{document}

\date{\textbf{Note: This version of the paper matches the version published in Monthly Notices of the Royal Astronomical Society. The definitive version is available at \href{http://mnras.oxfordjournals.org/content/422/3/2532}{www.blackwell-synergy.com}.}}

\volume{422}\pagerange{2532--2538} \pubyear{2012}

\maketitle

\label{firstpage}

\setcounter{page}{2532}

\begin{abstract}
In this paper, we show that the expansion history of the Universe in power-law cosmology essentially depends on two crucial parameters, namely the Hubble constant $H_{0}$ and deceleration parameter $q$. We find the constraints on these parameters from the latest $H(z)$ and SNe Ia data. At 1$\sigma$ level the constraints from $H(z)$ data are obtained as $q=-0.18_{-0.12}^{+0.12}$ and $H_{0}=68.43_{-2.80}^{+2.84}$ km s$^{-1}$ Mpc$^{-1}$ while the constraints from the SNe Ia data read as $q=-0.38_{-0.05}^{+0.05}$ and $H_{0}=69.18_{-0.54}^{+0.55}$ km s$^{-1}$ Mpc$^{-1}$. We also perform the joint test using $H(z)$ and SNe Ia data, which yields the constraints $q=-0.34_{-0.05}^{+0.05}$ and $H_{0}=68.93_{-0.52}^{+0.53}$ km s$^{-1}$ Mpc$^{-1}$. The estimates of $H_{0}$ are found to be in close agreement with some recent probes carried out in the literature. The analysis reveals that the observational data successfully describe the cosmic acceleration within the framework of power-law cosmology. We find that the power-law cosmology accommodates well the $H(z)$ and SNe Ia data.  We also test the power-law cosmology using the primordial nucleosynthesis, which yields the constraints $q\gtrsim 0.72$ and $H_{0}\lesssim 41.49$ km s$^{-1}$ Mpc$^{-1}$. These constraints are found to be inconsistent with the ones derived from the $H(z)$ and SNe Ia data. We carry out the statefinder analysis, and find that the power-law cosmological models approach the standard $\Lambda$CDM model as $q\rightarrow -1$. Finally, we conclude that despite having several good features power-law cosmology is not a complete package for the cosmological purposes.
\end{abstract}


\section{Introduction}

Power-law cosmology finds a reasonable place in the literature to
address some common problems (e.g., age problem, flatness problem, horizon problem etc.) associated with the Standard Cold Dark Matter (SCDM) model based on Big Bang theory. In such a cosmology, the cosmological evolution is described
by the scale factor $a(t)\propto t^\alpha$, where $\alpha$ is a
constant. The viability of the model with $\alpha\geq1$ has been
explored in a series of articles in different contexts
(Lohiya \& Sethi 1999; Batra et al. 1999, 2000; Gehlaut et al. 2002, 2003; Dev et al. 2002, 2008; Sethi et al. 2005a, 2005b; Zhu et al. 2008). Observational constraints on phantom
power-law cosmology are discussed by Kaeonikhom, Gumjudpai \& Saridakis (2011). The
motivation for such an endeavor is followed by a number of
considerations. For instance, power-law cosmological models with
$\alpha\geq1$ do not encounter the horizon problem at all (Sethi, Dev \& Jain 2005). These models do not witness the flatness
problem since the matter density is not constrained by the scale
factor. In these models, age of the Universe turns out at least
fifty percent greater than the age predicted by the SCDM
model. This bridges
the gap between the age of the Universe and the age estimates of
globular clusters and high-z redshift galaxies, and thus the age
problem is alleviated. Sethi, Dev \& Jain (2005)  showed
that an open linear coasting cosmological model constrained with
type Ia supernovae (SNe Ia) “gold” sample and ages of old
quasars accommodates a very old high-redshift quasar, which the
SCDM model fails to do. The linear coasting cosmology
is found to be consistent with the gravitational lensing statistics
(Dev et al. 2002) and the primordial nucleosynthesis
(Batra et al. 1999, 2000). Kaplinghat et al. (1999) found that the power-law cosmological models which succeed in primordial nucleosythesis are in conflict with the constraints from Hubble expansion rates and SNe Ia magnitude redshift relations.

In any cosmological model, the Hubble constant $H_{0}$ and
deceleration parameter $q$ play an important role is describing the
nature of evolution of the Universe. The former one tells us the
expansion rate of the Universe today while the latter one
characterizes the accelerating ($q<0$) or decelerating ($q>0$)
nature of the Universe. In the recent past, there have been numerous
attempts to estimate the value of $H_{0}$. Freedman et al. (2001) used the Hubble Space Telescope (HST) observations of
Cepheid variable to estimate a value of $H_{0}=72 \pm 8$ km s$^{-1}$ Mpc$^{-1}$. An observational determination of the Hubble constant obtained by
Suyu et al. (2010) based upon measurements of gravitational lensing
by using the HST yielded a value of $H_{0}= 69.7_{-5.0}^{+4.9}$
km s$^{-1}$ Mpc$^{-1}$. WMAP seven-year results gave an estimate of $H_{0}=
71.0\pm 2.5$ km s$^{-1}$ Mpc$^{-1}$ based on WMAP data alone, and an estimate of
$H_{0} = 70.4_{-1.4}^{+1.3}$ km s$^{-1}$ Mpc$^{-1}$ based on WMAP data with
Gaussian priors corresponding to earlier estimates from other
studies (Jarosik et al. 2011). A recent estimate of the Hubble constant,
which used a new infrared camera on the HST to measure the distance
and redshift for a collection of astronomical objects, gives a value
of $H_{0} = 73.8 \pm 2.4$ km s$^{-1}$ Mpc$^{-1}$ (Riess et al. 2011). An alternative
probe using data from galactic clusters gave a value of $H_{0}= 67.0
\pm 3.2$ km s$^{-1}$ Mpc$^{-1}$ (Beutler et al. 2011).

In this paper, we show that the power-law cosmology essentially
depends on the parameters $q$ and $H_{0}$. We intend to find the
observational constraints on the power-law cosmology parameters
using the recent observational data from $H(z)$ and supernova
observations. We also intend to test the power-law cosmology with primordial nucleosynthesis. The paper is organized as follows: The basic equations
of power-law cosmology are introduced in Section 2. Section 3 deals
with the constraints on the parameters $q$ and $H_{0}$ from the
latest $H(z)$ data while Section 4 is devoted to find the
constraints using Union2 compilation of 557 SNe Ia data points. In
Section 5, we perform the joint test using $H(z)$ and SNe Ia data. Section 6 deals with the constraints on power-law cosmology from primordial nucleosynthesis while Section 7 is devoted to study the statefinders in power-law cosmology.
In the last Section, we summarize the main results of
the paper.

\section{Basic Equations in Power-law Cosmology}
We study a general class of power-law cosmology described by the
dimensionless scale factor
 \begin{equation}\label{2}
a(t)=a_{0}\left(\frac{t}{t_{0}}\right)^\alpha,
\end{equation}
where $t_{0}$ is present age of the universe, $a_{0}$ is the value
of $a$ today and $\alpha$ is a dimensionless positive parameter.
Hereafter, the subscript $0$ denotes the present-day value of the
parameter under consideration.

The deceleration parameter $q$, which characterizes accelerating
($q<0$) or decelerating ($q>0$) nature of the Universe, reads as
\begin{equation}\label{3}
q=-\frac{a\ddot{a}}{\dot{a}^2}=\frac{1}{\alpha}-1,
\end{equation}
where an over dot denotes derivative with respect to cosmic time
$t$. The positivity of $\alpha$ leads to $q>-1$.

The cosmic scale factor in terms of the deceleration parameter may
be written as
\begin{equation}\label{4}
a(t)=a_{0}\left(\frac{t}{t_{0}}\right)^{1/(1+q)}.
\end{equation}

We observe that $q>-1$ is the condition for expanding Universe in
the power-law cosmological model.

The expansion history of the Universe is described by the Hubble
parameter,
\begin{equation}\label{5}
H(t)=\frac{\dot{a}}{a}=\left(\frac{1}{1+q}\right)\frac{1}{t},
\end{equation}
while the present expansion rate of the Universe is given by
$H_{0}=\frac{1}{(1+q)t_{0}}$.

The scale factor $a$ and the redshift $z$ are connected through the
relation $a=a_{0}(1+z)^{-1}$. Therefore, the Hubble parameter in
terms of the  redshift may be expressed as
\begin{equation}\label{6}
H(z)=H_{0}(1+z)^{1+q}.
\end{equation}

This shows that the expansion history of the Universe in power-law
cosmology depends on the parameters $H_{0}$ and $q$. 

It may be noted that the above model is well motivated in the literature as mentioned in the previous section. However the focus has been on one parameter namely $\alpha$ (or $q$). But here we find the observational constraints on both parameters $H_{0}$ and $q$ by subjecting the power-law cosmological model to the latest data from $H(z)$ and SNe Ia observations. We also constrain $H_{0}$ and $q$ using primordial nucleosynthesis scenario.

\section{Constraints from observational $H(z)$ data }
Simon, Verde \& Jimenez (2005)  determined nine $H(z)$ data
points in the range $0\leq z\leq 1.8$ by using the differential ages
of passively evolving galaxies determined from the Gemini Deep Deep
Survey and archival data. Recently, $H(z)$ data at 11
different redshifts based on the differential ages of red- envelope
galaxies were reported by Stern et al. (2010) while 3 more $H(z)$
data points were obtained by Gaztanaga, Cabre \& Hui (2009). The newly
$H(z)$ data points have been used to constrain parameters of various
cosmological models (Yang \& Zhang 2010; Cao, Zhu \& Liang 2011; Chen \& Ratra 2011; Paul, Thakur \& Ghose 2010, 2011).
Here, we use 13 observational $H(z)$ data points given in
Table 1 of the paper by Chen \& Ratra (2011) and the one at $z=0$ estimated in the work by Riess et al. (2011). For this sake, we define the $\chi^2$ as

\begin{equation}\label{7}
\chi^2_{H}(q,H_{0})=\sum_{i=1}^{14}\left[\frac{H(z_{i},q,H_{0})-H_{obs}(z_{i})}{\sigma_{i}}\right]^{2}.
\end{equation}

\begin{figure}
\psfrag{q}[b][b]{$q$}
\psfrag{h}[b][b]{$H_{0}$}
\centering
\includegraphics[width=8cm]{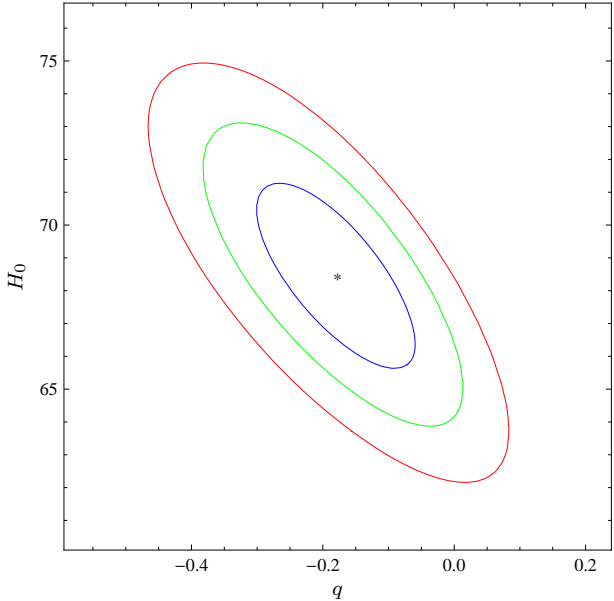}
\caption{The likelihood contours at 68.3\% (inner
contour), 95.4\% (middle contour) and 99.73\% (outer contour)
confidence levels  around the best fit values point $(-0.18,68.43)$
(shown by star symbol) in the $q-H_{0}$ plane obtained by fitting
power-law cosmological model with $H(z)$ data.} \label{fig1}
\end{figure}

The model has two free parameters namely $q$ and $H_{0}$. We perform
a grid search in the entire parametric space ($q>-1$ and
$H_{0}\geq0$) to find the best fit model. We find that the best fit
values of the parameters are $q=-0.18$ and $H_{0}=68.43$ together
with $\chi^2_{\nu}=1.49$, where $\chi^2_{\nu}=\chi^2_{min}$/(degree
of freedom). Here and in what follows $H_{0}$ is in the units of km s$^{-1}$ Mpc$^{-1}$. The negative value of $q$ suggests that the power-law
cosmological model fitted with the newly obtained $H(z)$ data
confirms the accelerating nature of the present-day Universe. The
likelihood contours at 68.3\% (inner contour), 95.4\% (middle
contour) and 99.73\% (outer contour) confidence levels around the
best fit values point $(-0.18,68.43)$ (represented by star symbol)
in the $q-H_{0}$ plane are shown in Fig.\ref{fig1}. The errors at
$1\sigma$ level are obtained as $q=-0.18_{-0.12}^{+0.12}$ and
$H_{0}=68.43_{-2.80}^{+2.84}$.

\section{Constraints from observational SNe Ia data }
The observations directly measure the apparent magnitude $m$ of a
supernova and its redshift $z$. The apparent magnitude $m$ of the
supernova is related to the luminosity distance $d_{L}$ of the
supernova through
\begin{equation}\label{8}
m=M+5\log_{10}\left(\frac{d_{L}}{1\;Mpc}\right)+25,
\end{equation}
where $M$ is the absolute magnitude, which is believed to be
constant for all SNe Ia.

It is convenient to work with Hubble free luminosity distance given
by
\begin{equation}\label{9}
D_{L}(z)=\frac{H_{0}}{c}d_{L}(z).
\end{equation}

Now, Eq.(\ref{8}) can be written as
\begin{equation}\label{10}
m=M+5\log_{10}D_{L}(z)-5\log_{10}H_{0}+52.38\;\;.
\end{equation}

The distance modulus $\mu(z)$ is given by
\begin{equation}\label{10}
\mu(z)=m-M=5\log_{10}D_{L}(z)-5\log_{10}H_{0}+52.38\;\;.
\end{equation}

The Hubble free luminosity distance $D_{L}$, in the present case,
for a geometrically flat Universe reads as
\begin{equation}\label{11}
D_{L}(z)=(1+z)\int_{0}^{z}\frac{H_{0}}{H(z')}dz'=\frac{1}{q}[(1+z)-(1+z)^{1-q}].
\end{equation}

SNe Ia are always used as standard candles, and are believed to
provide strongest constraints on the cosmological parameters. In the
present analysis, we use recently released Union2 set of 557 SNe Ia
from Supernova Cosmology Project (Amanullah et al. (2010). In this case,
we define the $\chi^2$ as
\begin{equation}\label{7}
\chi^2_{SN}(q,H_{0})=\sum_{i=1}^{557}\left[\frac{\mu(z_{i},q,H_{0})-\mu_{obs}(z_{i})}{\sigma_{i}}\right]^{2}.
\end{equation}

\begin{figure}
\psfrag{q}[b][b]{$q$}
\psfrag{h}[b][b]{$H_{0}$}
\centering
\includegraphics[width=8cm]{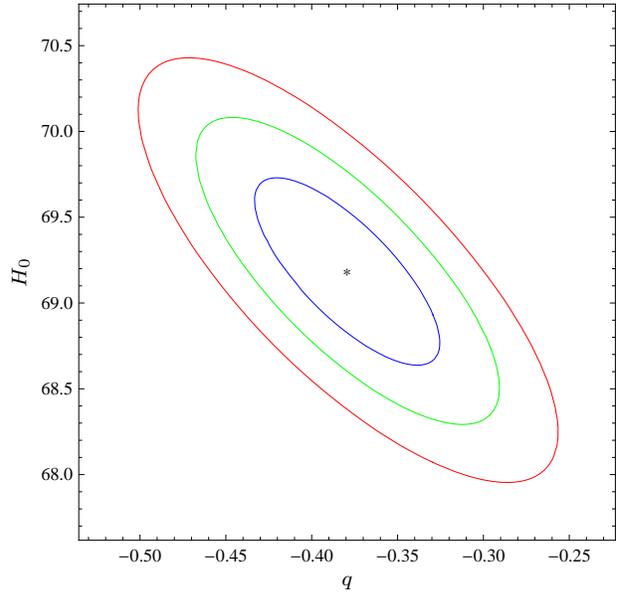}
\caption{The likelihood contours at 68.3\% (inner
contour), 95.4\% (middle contour) and 99.73\% (outer contour)
confidence levels around the best fit values point $(-0.38,69.18)$
(indicated by star symbol) in the $q-H_{0}$ plane obtained by
fitting power-law cosmological model with  SNe Ia data.}
\label{fig2}
\end{figure}

After performing a grid search in the entire parametric space
($q>-1$ and $H_{0}\geq0$), we find that the best fit values of the
parameters are $q=-0.38$ and $H_{0}=69.18$ together with
$\chi^2_{\nu}=0.99$. Again, the power-law cosmological model fitted
with the 557 SNe Ia data confirms the cosmic acceleration with
$q=-0.38$. The likelihood contours at 68.3\% (inner contour), 95.4\%
(middle contour) and 99.73\% (outer contour) confidence levels
around the best fit values point $(-0.38,69.18)$ (indicated by star
symbol) in the $q-H_{0}$ plane are shown in Fig.\ref{fig2}. The
errors at $1\sigma$ level are derived as $q=-0.38_{-0.05}^{+0.05}$
and $H_{0}=69.18_{-0.54}^{+0.55}$.

\section{Best-fitting model from the joint test: \boldmath $H(z)$+SNe Ia data }
In order to obtain tighter constraints on the model parameters and
to avoid degeneracy in the observational data, we combine $H(z)$ and
SNe Ia data. Since $H(z)$ and SNe Ia data are obtained from
independent cosmological probes, the total likelihood is considered
to be the product of separate likelihoods of the two probes.
Therefore, we define
\begin{equation}\label{7}
\chi^2_{total}=\chi^2_{H}+\chi^2_{SN}.
\end{equation}

\begin{figure}
\psfrag{q}[b][b]{$q$}
\psfrag{h}[b][b]{$H_{0}$}
\centering
\includegraphics[width=8cm]{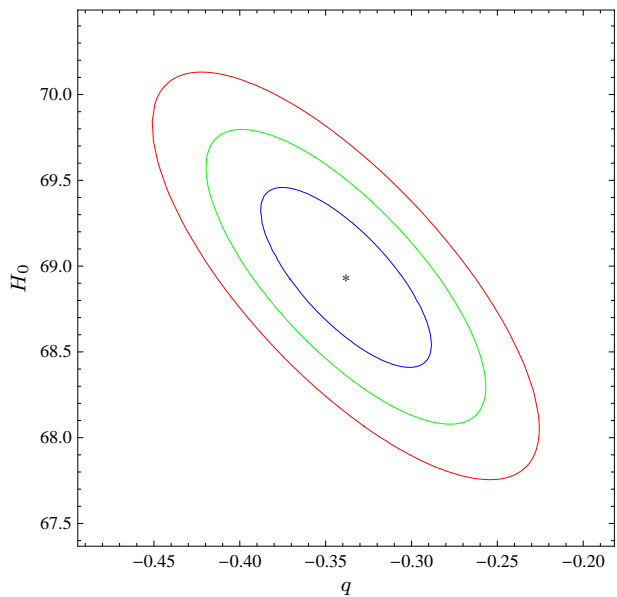}
\caption{The likelihood contours at 68.3\% (inner
contour), 95.4\% (middle contour) and 99.73\% (outer contour)
confidence levels around the best fit values point $(-0.34,68.93)$
(shown by star symbol) in the $q-H_{0}$ plane obtained by fitting
power-law cosmological model with $H(z)$+SNe Ia data.} \label{fig3}
\end{figure}

In the joint analysis, we find that the best fit values of the
parameters are $q=-0.34$ and $H_{0}=68.93$ together with
$\chi^2_{\nu}=1.01$. The joint test also confirms the cosmic
acceleration with $q=-0.34$.   The likelihood contours at 68.3\%
(inner contour) and 95.4\% (outer contour) confidence levels around
the best fit values point $(-0.34,68.93)$ (indicated by star symbol)
in the $q-H_{0}$ plane are shown in Fig.\ref{fig3}. The errors at
$1\sigma$ level read as $q=-0.34_{-0.05}^{+0.05}$ and
$H_{0}=68.93_{-0.52}^{+0.53}$.

\begin{figure*}
\psfrag{q}[b][b]{$z$}
\psfrag{h}[b][b]{$H(z)$}
\centering
\includegraphics[width=13cm]{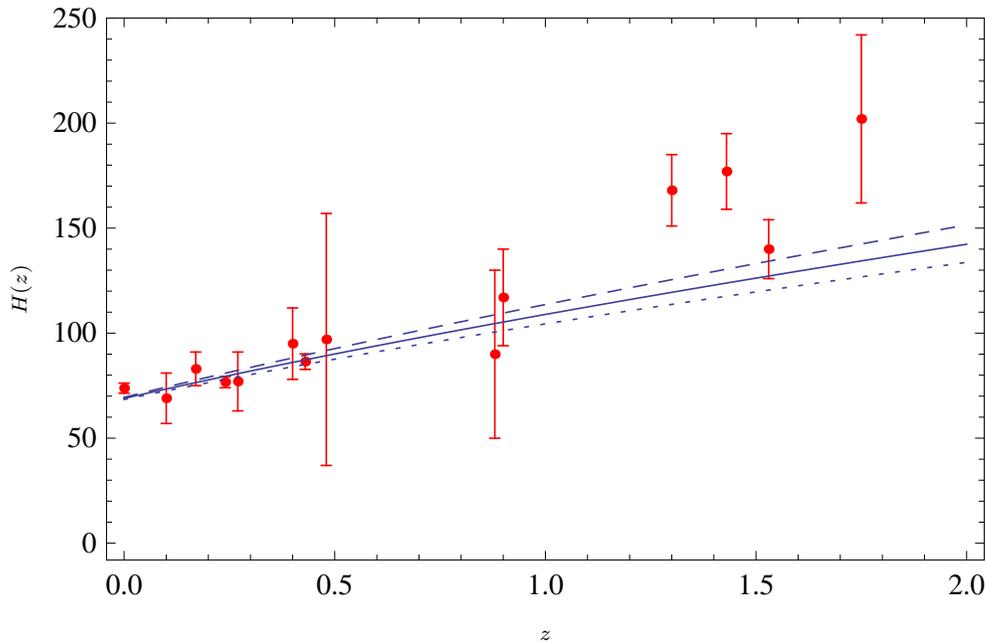}
\caption{The observational 15 $H(z)$ data points are
shown with error bars (red color online). Variation of best fit
model $H(z)$ curve (solid) based on $H(z)$+SNe Ia data is shown vs
$z$. The dashed curve corresponds to the maximum values of $H(z)$ in
the 1$\sigma$ region $68.41 \leq
H_{0}\leq 69.46, -0.39\leq q \leq
-0.29$ while the dotted curve corresponds to the minimum values of
$H(z)$ in the same region. We observe that the best fit model fits
well to the observational data points of $H(z)$ especially at
redshits $z<1$.} \label{fig4}
\end{figure*}

\begin{figure*}
\psfrag{q}[b][b]{$z$}
\psfrag{h}[b][b]{$\mu(z)$}
\centering
\includegraphics[width=13cm]{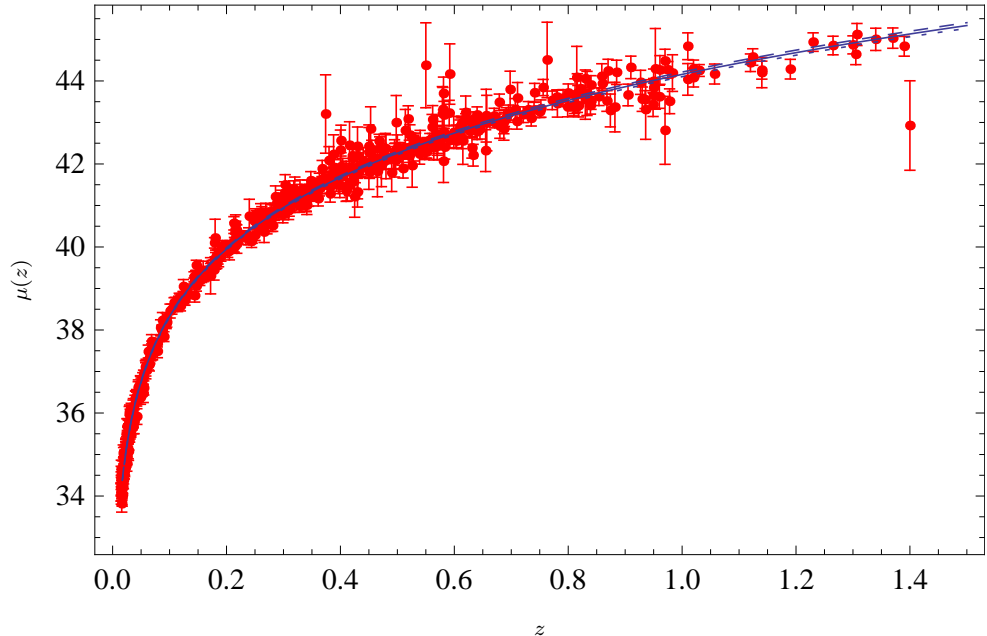}
\caption{ The observational 557 SNe Ia data points are
shown with error bars (red color online). Best fit model distance
modulus $\mu(z)$ curve (solid) based on $H(z)$+SNe Ia data is shown
vs $z$. The dashed curve and the dotted curve respectively
correspond to the maximum and minimum values of $\mu(z)$ in the
1$\sigma$ region $68.41 \leq
H_{0}\leq 69.46, -0.39\leq q \leq
-0.29$. We observe that the best fit model is in excellent agreement
with the observational data points of SNe Ia at all redshifts.}
\label{fig5}
\end{figure*}

Fig.\ref{fig4} demonstrates the comparison of the best fit
cosmological model obtained from the joint test with the
observational $H(z)$ data in the 1$\sigma$ region $68.41 \leq
H_{0}\leq 69.46 , -0.39\leq q \leq -0.29$. We observe that the model
fits well to the observational 15 $H(z)$ points shown with error
bars, especially at redshifts $z<1$.

In Fig.\ref{fig5}, the comparison of the derived best fit model
based on $H(z)$+SNe Ia with the observational 557 SNe Ia data points
(shown with error bars) of Union2 compilation in the 1$\sigma$
region $68.41 \leq H_{0}\leq 69.46 , -0.39\leq q \leq -0.29$ is
illustrated. We see that the model is in excellent agreement with
the observational SNe Ia data.

\section{Constraints from primordial nucleosynthesis}
Before we find constraints on power-law cosmology parameters from primordial big bang nucleosynthesis (BBN), it is helpful to reproduce the brief review of the BBN in standard model given by Kaplinghat et al. (1999). The neutron-proton ratio $n/p=\exp(-Q/T)$, where $Q=1.29$ MeV is the neutron-proton mass difference, at high temperatures $T\gtrsim 1 $ MeV is maintained by charged-current weak interactions among neutrons, protons, electrons, positrons and neutrinos. When the Universe is of order 1 s old, $T\lesssim$ 1 MeV, the $n/p$ ratio ``freezes out" due to the inequilibrium of weak interactions and free neutron decay with a lifetime of 887 s. At this stage the deuterium (D) produced
due to collision of neutrons and protons is rapidly photodissociated by the cosmic background photons. This causes very low abundance of D and thus heavier nuclei are not formed at this epoch. Thus nucleosynthesis is delayed by this ``photodissociation bottleneck". However, when the universe is $\sim$ 3 minutes old, the temperature falls below $\sim$ 80 keV, the deuterium bottleneck is broken. At this stage the nuclear reactions quickly burn out the remaining free neutrons into $^4$He and leave trace amounts of D, $^3$He and $^7$Li (Walker et al. 1991). A viable cosmological model must mimic the above scenario for proper synthesis of the light elements in the early Universe. In the following, we test the power-law cosmological model for primordial nucleosynthesis.

In power-law cosmology the scale factor $a(t)$ and the cosmic microwave background temperature $T(t)$ are related through the relation:
\begin{equation}
\frac{a}{a_{0}}=\frac{T_{0}}{\beta T}=\left(\frac{t}{t_{0}}\right)^\frac{1}{1+q},
\end{equation}
where $\beta$ stands for any non-adiabatic expansion due to entropy production. In standard cosmology, the instantaneous $e^{\pm}$ annihilation is assumed at $T=m_{e}$. The heating due to this annihilation is accounted by $\beta$ where $\beta=1$ for $T<m_{e}$ while $\beta=(11/4)^{1/3}$ for $T>m_{e}$. 

In order to find constraints from primordial nucleosythesis on power-law cosmology parameters, we utilize $t_{0}\approx 13.7$ Gyr (age of the Universe) estimated by Komatsu et al. (2011) on the basis of 7 year data from WMAP and astrophysical data from other sources. Further, the model is assumed to have current temperature $T_{0}=2.728$ K. The primordial nucleosynthesis requires that $t\lesssim 887$ s when $T\approx 80$ keV. This puts the following constraints on $q$ and $H_{0}$:\\

We observe that primordial nucleosynthesis requires decelerating expansion ($q>0$) of the Universe within the framework of power-law cosmology. This is contrary to outcome of accelerated expansion ($q<0$) of the Universe obtained in earlier sections by using latest data sets of $H(z)$ and SNe Ia observations. Also the primordial nucleosynthesis forces the Hubble constant $H_{0}\lesssim 41.49$, which is much smaller than its values estimated in earlier sections by using the updated data sets of $H(z)$ and SNe Ia observations. This shows that the power-law cosmological models which succeed in mimicking the nucleosynthesis scenario are in conflict with the constraints obtained on power-law cosmology parameters by using the latest observational data from $H(z)$ and SNe Ia. A similar conclusion on power-law cosmologies was drawn by Kaplinghat et al. (1999). However, in the case at hand, it has been done with updated observational data sets and the constraints have been obtained on both the parameters $q$ and $H_{0}$. 

\section{Constraints on statefinders in power-law cosmology}

\begin{figure*}\centering
{\psfrag{r1}[b][b]{$r$}
\psfrag{L}[b][b]{\footnotesize{$\Lambda$CDM}}
\psfrag{Acceleration Zone}[b][b]{\footnotesize{Acceleration Zone}}
\psfrag{s1}[b][b]{$s$}
\includegraphics[width=8 cm]{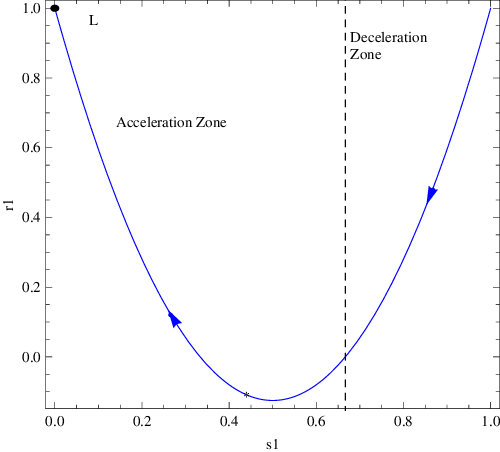}}
\hspace{0.5cm}
{
\psfrag{r1}[b][b]{$r$}
\psfrag{q1}[b][b]{$q$}
\psfrag{L}[b][b]{\footnotesize{dS}}
\includegraphics[width=8 cm]{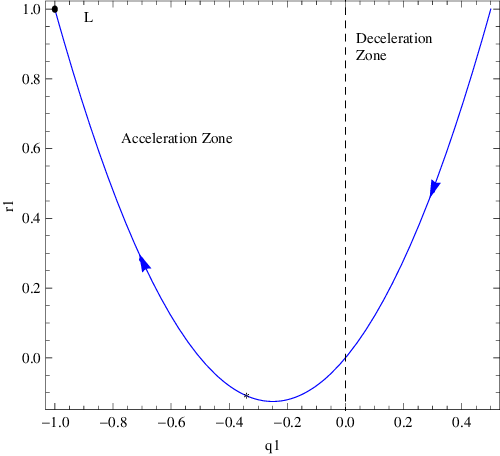}}
\caption{(a) Variation of $r$ versus $s$. The black
dot represents the statefinder pair $(s, r) = (0,1)$ or equivalently
location of flat $\Lambda$CDM model in the $s-r$ plane, and star
symbol on the $s-r$ curve shows the position of statefinder
parameters in the best fit model based on $H(z)$+SNe Ia
data. (b) Variation of $r$ versus $q$. The black dot
represents the location of de Sitter (dS) point $(q, r) = (-1, 1)$
in the $q-r$ plane, and star symbol on the $q-r$ curve shows the
position of $q$ and $r$ in the best fit model obtained from
$H(z)$+SNe Ia data. In both panels, the vertical dashed line
separates the acceleration and deceleration zones.  The arrows show
the direction of the evolution of trajectories as $q$ varies from
$0.5$ to $-1$. We observe that $\Lambda$CDM point $(0, 1)$ or
equivalently the dS point $(-1, 1)$ is an attractor  in power-law
cosmology.} \label{fig6}
\end{figure*}

The Hubble parameter $H=\dot{a}/a$ and deceleration parameter $q=-\ddot{a}/aH^2$ are useful geometric parameters in cosmology, which describe  the expansion history of the Universe. For instance,  $H>0$ ($\dot{a}>0$) indicates an expanding Universe while $q<0$ ($\ddot{a}>0$) characterizes an accelerated expansion of the Universe. In order to explain accelerated expansion of the Universe, various dark energy models have been proposed in the literature. However, these models encounter degeneracy on the geometric parameters $H$ (involving first time derivative of scale factor) and $q$ (involving second time derivative of scale factor) at the present epoch (Malekjani \& Khadom-Mohamaddi 2012). Thus the geometric parameters $H$ and $q$ are not capable of discriminating between different dark energy models and a viable diagnostic tool is required for this purpose. Sahni et al. (2003) proposed a pair of parameters
$\left\{r,s\right\}$ called statefinders as a means of distinguishing between
different dark energy models. The statefinders
involve derivatives of scale factor upto third order and are defined as
\begin{center}$r=\frac{\dot{\ddot{a}}}{aH^{3}}\;\;\;\;\;\;$ and
$\;\;\;\;\;\;\;s=\frac{r-1}{3(q-1/2)}$.\end{center} 

The remarkable feature of statefinders is that these parameters depend on scale factor and its time derivatives, and hence are geometric in nature (Sahni 2002). Further, different dark energy models exhibit different evolutionary trajectories in the $s-r$ plane. Moreover, the well known flat $\Lambda$CDM model corresponds to the point $s=0$ and $r=1$ in the $s-r$ plane (Alam et al. 2003). These features of statefinders provide an opportunity to distinguish between different dark energy models. The statefinder diagnostic tool has been extensively used in the literature  as a means
to distinguish between different dark energy models 
(see Alam et al. 2003; Ali et al. 2010; Malekjani \& Khadom-Mohamaddi 2012 and references therein). In what follows, we test the power-law cosmology by statefinder diagnostic tool.

In power-law cosmology, the statefinders read as

\begin{center}$r=2q^2+q\;\;\;\;\;\;$ and
$\;\;\;\;\;\;\;s=\frac{2}{3}(q+1)$, where $q\neq \frac{1}{2}$.\end{center}

We immediately notice that $r=1$ and $s=0$ at $q=-1$. Thus the the power-law cosmology mimics the $\Lambda$CDM model at $q=-1$. We obtain the evolutionary $s-r$ and $q-r$ trajectories for the power-law cosmology as shown in Fig.
\ref{fig6} for the values of $q$ in the range $-1\leq q<0.5$ . The black dot  in the left panel (a) at $(s, r) = (0,1)$
represents the location of flat $\Lambda$CDM model while the the
black dot in the right panel (b) shows the location of the de Sitter
(dS) point  $(q, r) =(-1, 1)$. The star symbol in both panels
corresponds to the best fit model based on $H(z)$+SNe Ia. It is
interesting to note that the $\Lambda$CDM statefinder pair $(0, 1)$
or equivalently the dS point $(-1, 1)$ is an attractor in power-law
cosmology.  

Now we find the constraints on the statefinders of power-law cosmology from $H(z)$, SNe Ia and BBN observations. The $H(z)$ data constrain the statefinders as $r=-0.09_{-0.03}^{+0.04}$ and $s=0.58_{-0.12}^{+0.04}$ while the SNe Ia data constraints on statefinders are $r=-0.09_{-0.02}^{+0.03}$ and $s=0.41_{-0.03}^{+0.03}$. The joint test of $H(z)$ and SNe Ia data puts the following constraints on statefinders: $r=-0.11_{-0.01}^{+0.02}$ and $s=0.44_{-0.03}^{+0.03}$. The errors in the above values are at $1\sigma$ level. 
The primordial nucleosynthesis restricts $r\gtrsim 1.76$ and $s\gtrsim 1.15$. We observe that the best fit values of statefinders $r$ and $s$ predicted by observational $H(z)$ and SNe Ia data are in conflict with the ones estimated by BBN, as expected.

\section{Summary}
In this paper, we have found the bounds on the parameters $H_{0}$
and $q$ of the power-law cosmology. The numerical results are
summarized in Table 1.

\begin{table*}
\textbf{Table 1.} Summary of the numerical results\\
\begin{tabular}{l c c c c c}
\hline
Data/Source & $q$ & $H_{0}$  & $\chi^2_{\nu}$& $r$& $s$\\
\hline
$H(z)$ & $-0.18_{-0.12}^{+0.12}$ & $68.43_{-2.80}^{+2.84}$ & $1.49$ & $-0.09_{-0.03}^{+0.04}$ & $0.58_{-0.12}^{+0.04}$ \\\\
  SNe Ia & $-0.38_{-0.05}^{+0.05}$ &$69.18_{-0.54}^{+0.55}$& $0.99$& $-0.09_{-0.02}^{+0.03}$ & $0.41_{-0.03}^{+0.03}$   \\\\
 $H(z)+$SNe Ia& $-0.34_{-0.05}^{+0.05}$ &  $68.93_{-0.52}^{+0.53}$& $1.01$& $-0.11_{-0.01}^{+0.02}$ & $0.44_{-0.03}^{+0.03}$   \\\\
$BBN$&$\gtrsim 0.72$ &  $\lesssim 41.49$& $--$& $\gtrsim 1.76$ & $\gtrsim 1.15$   \\
 
\hline
\end{tabular}
\label{table:nonlin}
\end{table*}

Some key observations are as follows:

\begin{itemize}
\item [(i)] The constraints on the deceleration parameter $q$ clearly indicate that the astronomical observations of $H(z)$ and SNe Ia predict the cosmic acceleration within the framework of power-law cosmology. Thus, power-law cosmological models are viable for describing the observed accelerating nature of Universe.

\item [(ii)] We see that the estimates of Hubble constant in power-law cosmology are in close agreement with independent investigations of $H_{0}$ carried out in literature (Freedman et al. 2001; Suyu et al. 2010; Jarosik et al. 2010; Riess et al. 2011; Beutler et al. 2011) as discussed in Section 1.

\item [(iii)] The derived best-fitting model fits well to the observational data points from $H(z)$ and SNe Ia observations (see Figs
\ref{fig4} and
\ref{fig5}).

\item [(iv)] The primordial nucleosynthesis demands a decelerating expansion of the Universe with smaller values of Hubble constant.

\item [(v)] The statefinder analysis shows that power-law cosmological models approach the standard $\Lambda$CDM model in future (see Fig.
\ref{fig6}) with varying values of $q$.

\end{itemize}

We see that the power-law cosmology turns out viable in the description of the acceleration of present-day Universe when subjected to recent observations of $H(z)$ and SNe Ia. Also the Hubble constant within the framework of power-law cosmology falls in the range of observations. However, the power-law cosmology fails to produce primordial nucleosytheis with the values of  $q$ and $H_{0}$ estimated from observational data of $H(z)$ and SNe Ia as discussed in Section 6. Moreover, because of the constant value of deceleration parameter $q$ in power-law cosmology, it fails to provide time or redshift based transition of the Universe from deceleration to acceleration. Thus, one has to use different values of $q$ for the description of Universe at different epochs. Finally, despite having several useful features, the power-law cosmology is not a complete package for cosmological purposes. 
\section*{Acknowledgements}
The author is thankful to M. Sami, Deepak Jain and Sanjay Jhingan for fruitful discussions. Thanks are also due to Bharat Ratra and \"{O}zg\"{u}r Akarsu for useful comments on the initial draft of the manuscript. The author would like to put forward his sincere thanks to the referee whose valuable comments have helped in improving the quality of this manuscript.

\section*{References}
Alam U., Sahni V., Saini T.D., Starobinsky A.A.,  2003, \href{http://onlinelibrary.wiley.com/doi/10.1046/j.1365-8711.2003.06871.x/abstract}{MNRAS}, 344, 1057\\
Ali A., Gannouji R., Sami M., Sen A.A., 2010, Phys. Rev. D, 81, 104029\\ 
Amanullah R. et. al, 2010, \href{http://iopscience.iop.org/0004-637X/716/1/712}{ApJ}, 716, 712 \\
Batra A., Sethi M., Lohiya D., 1999, Phys. Rev. D, 60, 108301\\
Batra A., Lohiya D., Mahajan S.,  Mukherjee A., 2000, Int. J. Mod. Phys. D, 9, 757\\
Beutler F. et al., 2011, \href{http://onlinelibrary.wiley.com/doi/10.1111/j.1365-2966.2011.19250.x/abstract}{MNRAS}, 416, 3017\\
Cao S., Zhu Z.-H., Liang N., 2011, A $\&$ A, 529, A61\\ 
Chen Y., Ratra, B., 2011, preprint  \href{http://arxiv.org/abs/1106.4294v2}
{(astro-ph/1011.4848)}\\
Dev A., Jain D., Lohiya D., 2008, preprint \href{http://arxiv.org/abs/0804.3491v1}
{(astro-ph/0804.3491)}\\
Dev A., Safanova M., Jain D., Lohiya D., 2002, Phys. Lett. B, 548, 12 \\
Freedman W.L. et al., 2001, ApJ, 553, 47\\ 
Gaztanaga E., Cabre A., Hui L., 2009, MNRAS, 399, 1663\\ 
Gehlaut S., Mukherjee A., Mahajan S., Lohiya D., 2002, Spacetime $\&$ Substance, 4, 14\\
Gehlaut S., Kumar P., Sethi G., Lohiya D., 2003, preprint \href{http://arxiv.org/abs/astro-ph/0306448v1}
{(astro-ph/0306448)}\\
Jarosik N. et al., 2010, \href{http://dx.doi.org/10.1088/0067-0049/192/2/14}{ApJS}, 192, 14\\ 
Kaeonikhom C., Gumjudpai B., Saridakis E.N., 2011, Phys. Lett. B, 695, 45\\ 
Kaplinghat M., Steigman G., Tkachev I., Walker, T.P., 1999, Phys. Rev. D, 59, 043514\\
Komatsu E. et al., 2011, \href{http://iopscience.iop.org/0067-0049/192/2/18}{ApJS}, 192, 18\\  
Lohiya D., Sethi M., 1999, Class. Quantum Grav., 16, 1545\\
Malekjani M., Khodam-Mohammadi A., 2012, preprint \href{http://arxiv.org/abs/1202.4154v1}
{(gr-qc/1202.4154)}\\
Paul B.C., Thakur P., Ghose S., 2010, \href{http://onlinelibrary.wiley.com/doi/10.1111/j.1365-2966.2010.16909.x/abstract}
{MNRAS}, 407, 415 \\
Paul B.C., Ghose S., Thakur P., 2011 \href{http://onlinelibrary.wiley.com/doi/10.1111/j.1365-2966.2010.18177.x/abstract}
{MNRAS}, 413, 686\\ 
Riess A.G. et al., 2011, \href{http://dx.doi.org/10.1088/0004-637X/730/2/119}{ApJ}, 730, 119\\
Sahni V., 2002, preprint \href{http://arxiv.org/abs/astro-ph/0211084}{(astro-ph/0211084)} \\ 
Sahni V., Saini T.D., Starobinsky A.A., Alam U., 2003, Pisma Zh. Eksp. Teor. Fiz., 77, 249\\ 
Sethi G., Kumar P., Pandey S., Lohiya D., 2005, Spacetime $\&$ Substance, 6, 31\\ 
Sethi G., Dev A., Jain D., 2005, Phys. Lett. B, 624, 135\\ 
Simon J., Verde, L., Jimenez, R., 2005, Phys. Rev. D, 71, 123001\\ 
Stern D. et al., 2010, JCAP, 02, 008\\ 
Suyu S.H. et al., 2010, \href{http://iopscience.iop.org/0004-637X/711/1/201/}{ApJ}, 711, 201\\
Walker T.P., Steigman G., Schramm K.A., Olive K.A., Kang H.-S., 1991, ApJ, 376, 51\\
Yang R.-J., Zhang S.N., 2010, \href{http://onlinelibrary.wiley.com/doi/10.1111/j.1365-2966.2010.17020.x/
abstract}
{MNRAS}, 407, 1835 \\ 
Zhu, Z.-H., Hu, M., Alcaniz, J.S., Liu, Y.-X., 2008, A $\&$ A, 483, 15

\end{document}